\documentclass[a4paper,11pt]{article}
\usepackage{pos}
\usepackage{lineno}

\usepackage{ptdr-definitions}
\usepackage{hepunits}
\usepackage{heppennames2}
\title{Evidence for top quark production in nucleus-nucleus collisions}
\newcommand{\rootsNN} {\ensuremath{\sqrt{\smash[b]{s_{_{\mathrm{NN}}}}}}\xspace}
\newcommand{\roots} {\ensuremath{\sqrt{\smash[b]{s}}}\xspace}
\newcommand{\PbPb}{\ensuremath{\mathrm{Pb}\mathrm{Pb}}\xspace}
\newcommand{\NN}{\ensuremath{\mathrm{N}\mathrm{N}}\xspace}
\newcommand{\pp}{\ensuremath{\Pp\Pp}\xspace}

\newcommand{\stt}{\ensuremath{\sigma_{\ttbar}}\xspace}

\newcommand{\RecoLumi}{\ensuremath{ 1.7\pm 0.1\nbinv}}
\newcommand{\empm}{\ensuremath{\Pe^\pm \Pgm^\mp}}
\newcommand{\mmpm}{\ensuremath{\Pgm^+ \Pgm^-}}
\newcommand{\empe}{\ensuremath{\Pe^+ \Pe^-}}
\newcommand{\eb}{\ensuremath{\varepsilon_{\cPqb}}}
\newcommand{\dQGP}{\ensuremath{\delta_{\text{QGP}}}}
\newcommand{\levone}{\ensuremath{2\ell_\text{OS}}}
\newcommand{\levtwo}{\ensuremath{2\ell_\text{OS}{}+{}\text{b-tags}}}
\providecommand{\PN}{\ensuremath{\mathrm{N}}\xspace}
\newcommand{\VV}{\ensuremath{\cmsSymbolFace{V}\cmsSymbolFace{V}}\xspace}
\newcommand{\mub}{\ensuremath{\,\mu\text{b}}\xspace}

\newcommand{\zvalexptwol}{\ensuremath{ 4.8}}
\newcommand{\zvalobstwol}{\ensuremath{ 3.8}}
\newcommand{\zvalexptwoljets}{\ensuremath{ 5.8}}
\newcommand{\zvalobstwoljets}{\ensuremath{ 4.0}}

\newcommand{\muobstwol}{\ensuremath{ 0.79^{+0.26}_{-0.23}}}

\newcommand{\muobstwoljets}{\ensuremath{ 0.63^{+0.22}_{-0.20}}}
\newcommand{\stttwol}{\ensuremath{2.54^{+0.84}_{-0.74}}}
\newcommand{\stttwoljets}{\ensuremath{2.03^{+0.71}_{-0.64}}}
\author*[a]{Georgios Konstantinos Krintiras}
\affiliation[a]{The University of Kansas\\ \href{https://gkrintir.web.cern.ch/}{cern.ch/gkrintir}}

\emailAdd{gkrintir@cern.ch}

\abstract{Using $1.7 \pm 0.1\,\nbinv$ of lead-lead ($A = 208$) collision data recorded by the CMS experiment at a nucleon-nucleon center-of-mass energy of 5.02 \TeV, we report evidence of top quark pair (\ttbar) production. The \ttbar cross section (\stt) is  extracted from likelihood fits to a multivariate discriminator using lepton kinematic variables in dilepton final states and two methods. One method relies on the leptonic information alone, and the second one exploits, in addition, the presence of bottom quarks. The measured \stt is \stttwol\ and \stttwoljets \mub in the two cases, respectively, consistent with predictions from perturbative quantum chromodynamics. We demonstrate, for the first time, that top quark decay products (leptonically decaying \PW bosons and bottom quarks) can be identified, irrespective of any possible final-state interactions with the quark-gluon plasma.}

\FullConference{%
  HardProbes2020\\
  1--6 June 2020\\
  Austin, Texas}

\begin{document}
\maketitle


Droplets of quark-gluon plasma (QGP), a state of strongly interacting quantum chromodynamics (QCD) matter, are produced in heavy nuclei high-energy collisions. A detailed study of the top (\PQt) quark decay products, \ie, bottom (\cPqb) quarks and \PW\ bosons~\cite{TANABASHI:2018OCA}, provides novel insights into the mechanisms of QGP-induced parton energy loss. On one hand, \cPqb quarks are ideally suited to serve as a ``standard candle'' of the amount of energy suppression for \cPqb quark jets~\cite{Chatrchyan:2013exa} emerging almost simultaneously with the heavy ion collisions~\cite{Baskakov:2015nxa}.
On the other hand, hadronically decaying \PW\ bosons are not immediately resolved by the medium, and can probe the QGP density evolution at different space-time scales~\cite{Apolinario:2017sob,Dainese:2016gch}. The length of the time delay can be constrained by selecting a range of reconstructed top quark transverse momentum (\pt). 

At the LHC, top quark cross sections are dominated by pair production (\ttbar) via gluon-gluon ({\cPg{}\cPg}) fusion, \ie, the partonic reaction $\cPg\cPg \to \ttbar+X$, and are
computable with high accuracy in perturbative quantum chromodynamics (QCD)~\cite{Czakon:2011xx,Czakon:2013goa,Catani:2019hip}. The top quark is thus a theoretically precise probe of the initial state, and in particular the gluon parton distribution function (PDF)~\cite{Sirunyan:2017ule}. Top quark pair production is characterized by final states~\cite{TANABASHI:2018OCA} comprising the decay products of the two \PW\ bosons and two \cPqb jets.

Although the feasibility of top quark studies with nuclear collisions was demonstrated recently~\cite{Sirunyan:2017xku},
they remained inaccessible in nucleus-nucleus collisions because of the amount of data accumulated so
far at the CERN LHC. At the end of 2018, LHC provided its four major experiments with lead-lead (\PbPb) collisions at a nucleon-nucleon (\NN) center-of-mass energy $\rootsNN = 5.02\,\TeV$ and an unprecedented collision rate.
The amount of data recorded by the CMS experiment~\cite{Chatrchyan:2008zzk} corresponds to an integrated luminosity of about \RecoLumi~\cite{CMS-PAS-LUM-17-002}, making the first-ever measurement of top quark  production in nucleus-nucleus collisions possible~\cite{Sirunyan:2020kov}.

\section{Methods}
\label{sec:Methods}

We first identify \ttbar signal events based on the leptonic $\PW^{\pm} \to \ell^{\pm}\PGn$ decays, with $\ell$ being either an electron (\Pe) or muon (\Pgm).
The resulting dilepton final states (\empm, \mmpm, and \empe) involve two energetic, oppositely charged ($\mathrm{OS}$) leptons that are isolated from nearby hadronic activity, momentum imbalance from the undetected neutrinos (\PGn), and two \cPqb\ jets. Two methods are
employed: (i) making use of the final-state
dilepton kinematic properties alone (``\levone''), and (ii) imposing extra requirements on the number of ``\PQb-tagged jets''  (``\levtwo''). Since the feasibility to reconstruct \cPqb jets in the QGP may be impacted by the sizeable suppression evidenced in data, the \levtwo method relies on a data-based estimate for the \cPqb\ jet identification (``tagging'') performance, measured in terms of the \cPqb\ jet identification efficiency \eb.
Finally, we extract the \ttbar cross section (\stt) from a combined maximum-likelihood fit to a multivariate discriminator using lepton kinematic variables, 
independently for the \levone\ and \levtwo\ methods.

The $\PN\PN \to \ttbar+X$ process ($\PN = \Pp,\Pn$) is simulated with the \MGvATNLO(v2.4.2)~~\cite{Alwall:2014hca} program at next-to-leading order (NLO) with the EPPS16 NLO nuclear PDF~\cite{ESKOLA:2016OHT}. Drell--Yan production of quark-antiquark annihilation into lepton-antilepton pairs 
through \cPZ boson or virtual-photon exchange (referred to as ``$\cPZ{}/\gamma^{*}$'') and \PW boson production with additional jets, are generated with the same program at NLO, corrected with scaling factors derived from data. Subdominant contributions from single top quark plus \PW{} boson events ($\PQt\PW$) and \PW{}\PW, \PW{}\cPZ, and \cPZ{}\cPZ\ production (collectively referred to as ``\VV'') are simulated with the \POWHEG~\cite{Alioli:2010xd} NLO generator. 

Gradient boosted decision trees (BDTs) are set up to maximally discriminate genuine leptons with high \pt\ between the \ttbar signal and the $\cPZ{}/\gamma^{*}$ background. The BDTs exploit kinematic properties of the leading- and subleading-\pt leptons (referred to as ``$\ell_1$'' and ``$\ell_2$'', respectively). The maximum-likelihood fits are performed on binned BDT distributions, simultaneously in the \empm, \mmpm, and \empe\ final states, and accounting for all sources of uncertainty, \ie, statistical and systematic, and their correlations. The simultaneous analysis of these events allows the measurement of the \ttbar signal strength $\mu$, defined as
the ratio of the observed \stt\ to the expectation from theory. The best-fit value of $\mu$ and its uncertainty $\Delta\mu$ (corresponding to a 68\% confidence level) are extracted separately with the \levone\ and \levtwo\ methods. For the latter, we correlate the number of \ttbar signal events in the \cPqb{}-tagged jet categories based on multinomial probabilities,
using \eb\ and a parameter (\dQGP) accounting for medium-induced suppression of \eb. We allow \eb\ being different for the two \cPqb\ jets, i.e.,
$\eb \to \eb^*=(1-\dQGP) \times \eb$, motivated by a path-length dependence of the parton energy loss.

\section{Results}
\label{sec:Results}

\begin{figure}[!htp]
\centering
\includegraphics[width=0.32\textwidth]{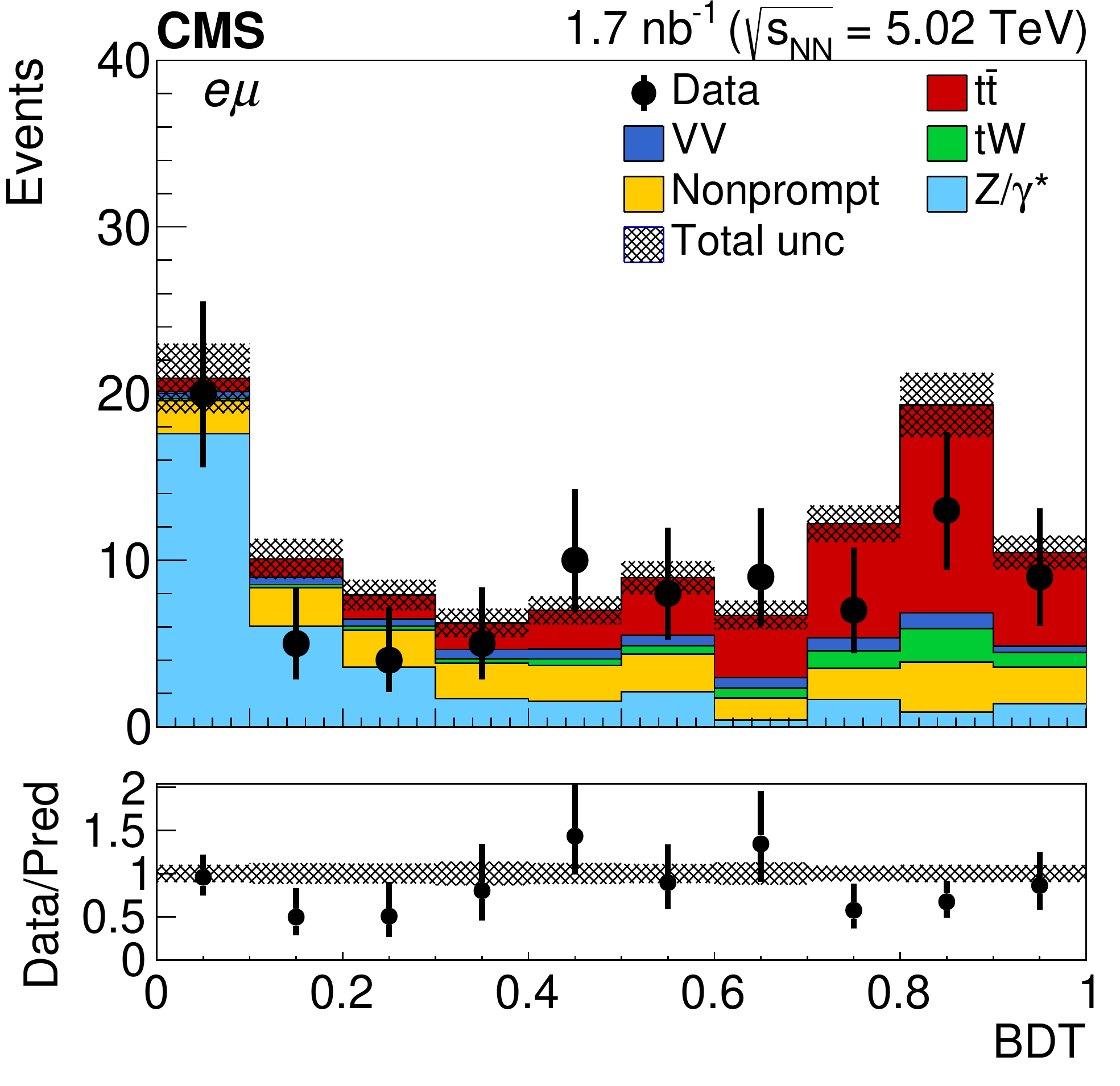}
\includegraphics[width=0.32\textwidth]{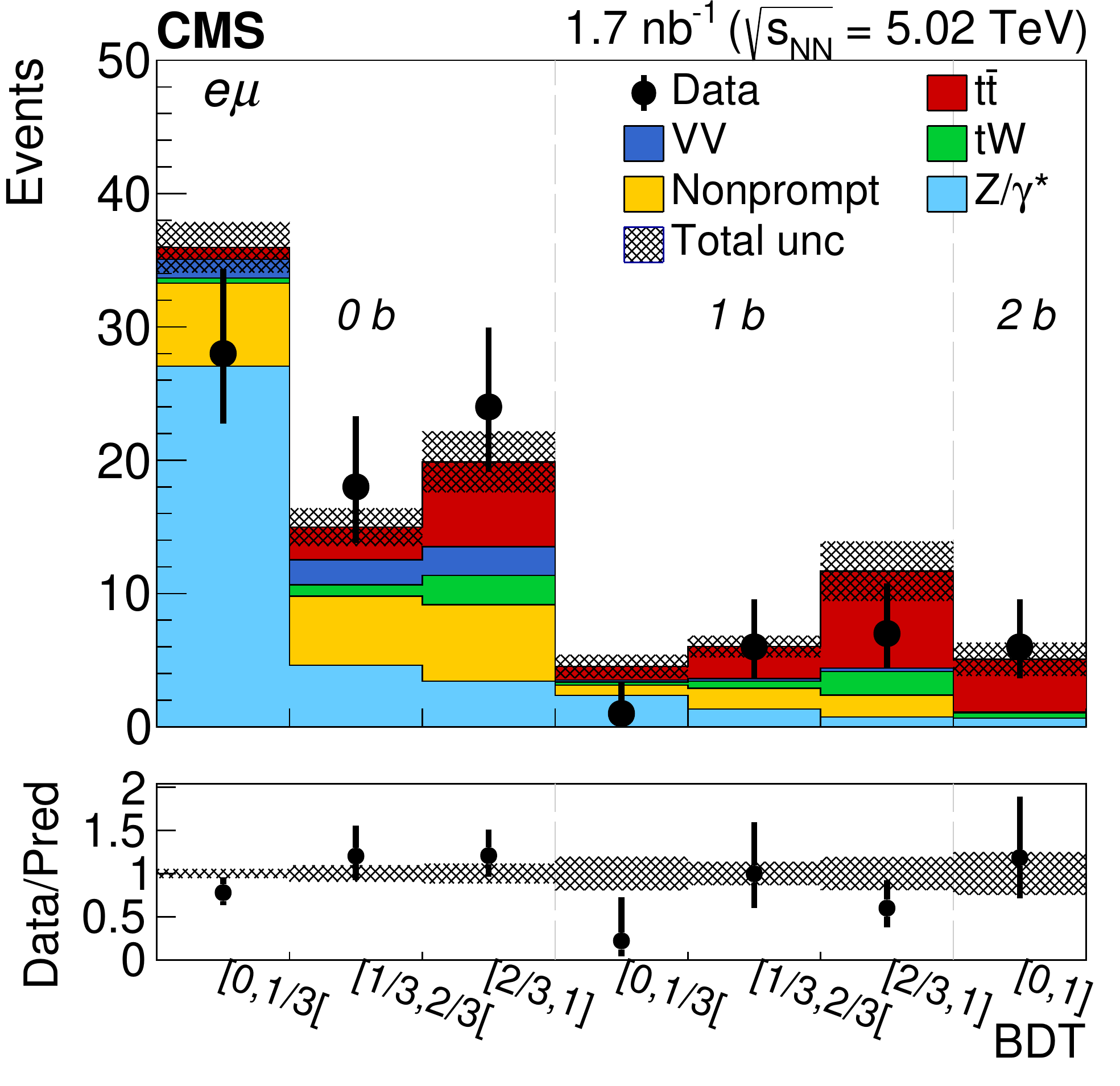}
\caption{
  Postfit predicted (histograms) and observed (points) binned BDT (left, middle) and sphericity (right) distributions in the \empm\ final state with the \levone\ (left, right) and \levtwo\ (middle) methods.
  The comparison between the \ttbar\ signal and the background-subtracted data is shown for the postfit distributions as inset panel (right).
  The vertical bars on the points represent the statistical uncertainty in data (68\% Clopper-Pearson intervals).
  The hatched regions show the postfit uncertainty in the sum of \ttbar\ signal and background. 
  The lower panels display the ratio of the observed data to the predictions, including the \ttbar signal, with bars and bands, respectively,
  representing the statistical and total uncertainties in the prediction~\cite{Sirunyan:2020kov}.
    }
\label{fig:results}
\end{figure}

The ``postfit predicted'' (\ie, the \ttbar\ signal and various sources of background are adjusted according to the fit procedure of Section~\ref{sec:Methods}) and observed BDT distributions are shown in Fig.~\ref{fig:results} in the \ttbar{}-enriched \empm\ final state with the \levone\ ($\mu=\muobstwol$, left) and \levtwo\ ($\mu=\muobstwoljets$, right) methods. 
We found the classifier to separate well the \ttbar signal from the \cPZ/$\gamma^{*}$ background in the \mmpm\ and \empe\ final states, 
enhancing our confidence about its applicability to the \empm\ final state. The inclusive \stt\ is then obtained multiplying 
the best fit $\mu$ value by the theoretical expectation. Accounting for the acceptance corrections, we measure
\stt to be \stttwol\ and \stttwoljets \mub in the combined \empm, \mmpm, and \empe\ final states, with a relative total uncertainty of 32 and 34\% in the \levone\ and \levtwo\ methods, respectively.

The compatibility of the data with the background-only hypothesis is evaluated using
a profile-likelihood ratio as a test statistic, including all sources of systematic uncertainty.
The probability for the background to mimic an excess of events larger than that observed in data is quantified using $p$ values, expressed in terms of Gaussian tail probabilities and given in units of standard deviation ($\sigma$).
The background-only hypothesis is excluded with observed (expected) statistical significance of \zvalobstwol\ (\zvalexptwol) and \zvalobstwoljets\ (\zvalexptwoljets)\,$\sigma$
with the \levone\ and \levtwo\ methods, respectively. 

\begin{figure}[!htb]
\centering
\includegraphics[width=0.50\textwidth]{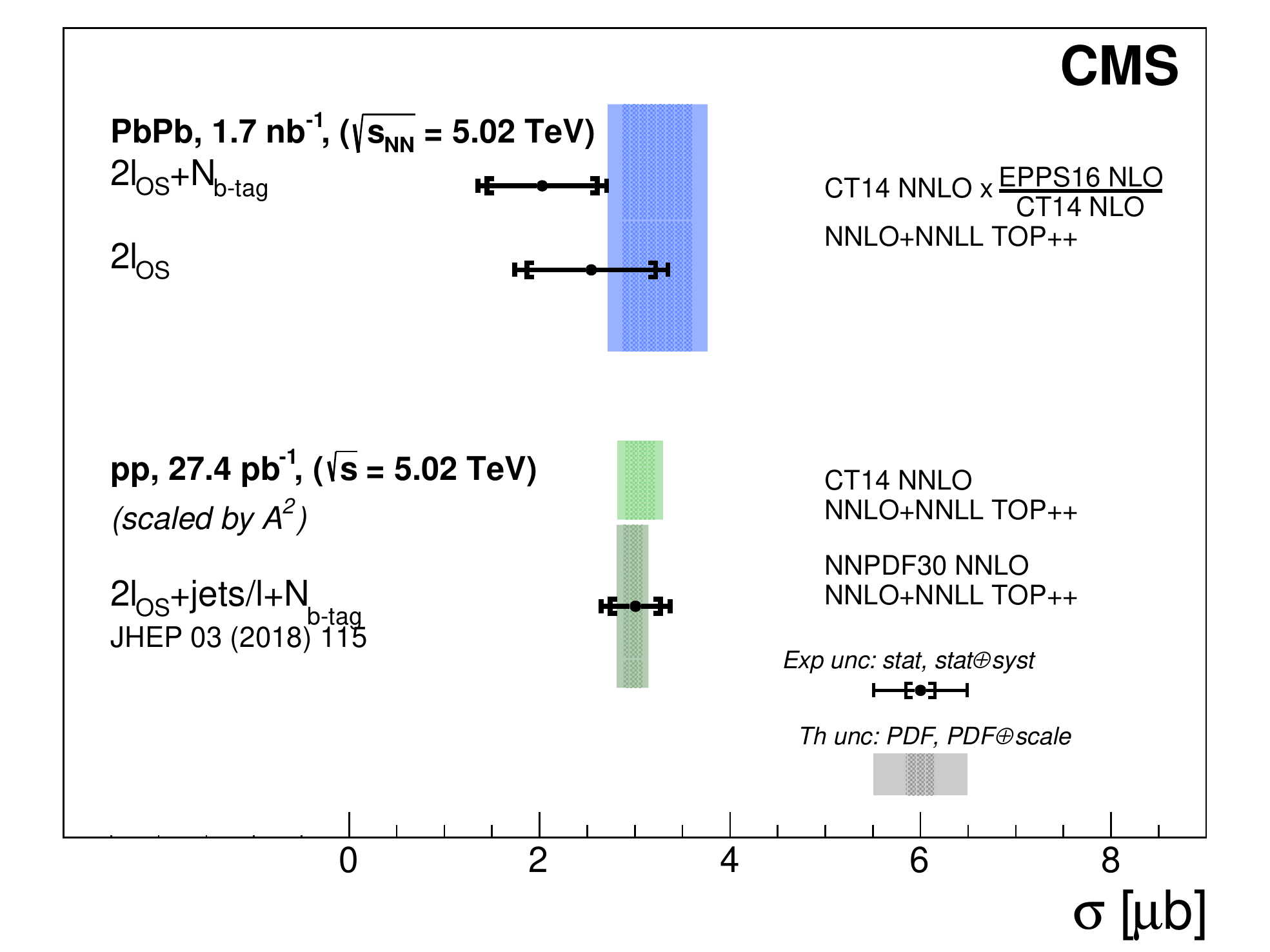}
\caption{The inclusive \ttbar\ cross section measured in the combined \empm, \mmpm, and \empe final states in \PbPb\ collisions (divided by the mass number squared, $A^2$),
  compared to predictions~\cite{Czakon:2013goa}, and \pp\ results at $\rootsNN=5.02$ \TeV~\cite{Sirunyan:2017ule}.
  The total experimental error bars (theoretical error bands) include statistical and systematic (PDF and QCD scale) uncertainties added in quadrature~\cite{Sirunyan:2020kov}.
  }
\label{fig:summary}
\end{figure}

Figure~\ref{fig:summary} presents the measured \stt, including the measurement at $\roots=5.02$ \TeV~\cite{Sirunyan:2017ule} in proton-proton (\pp) collisions,
and compared to predictions from perturbative QCD~\cite{Czakon:2013goa}.

\section{Summary}
\label{sec:Summary}

In summary, the top pair production cross section (\stt) is measured for the first time in nucleus-nucleus collisions, 
using lead-lead collision at $\rootsNN = 5.02\,\TeV$ with an integrated luminosity of \RecoLumi. The measurement makes use of at least one pair of oppositely charged electrons or muons, and, separately, jets originating from the bottom quarks. The extracted \stt\ is \stttwol\ and \stttwoljets \mub, respectively, consistent with the expectations from scaled proton-proton data as well as perturbative quantum chromodynamics calculations. The observed statistical significance of the \ttbar\ signal against the background-only hypothesis is \zvalobstwol\ and \zvalobstwoljets\ standard deviations in the two cases, respectively.

The measurement provides a novel tool for probing nuclear parton distribution functions, the mechanisms of parton energy loss, and the medium opacity at different space-time scales in nuclear interactions with increased LHC heavy ion luminosities or future higher-energy colliders.

\bibliographystyle{auto_generated}
\bibliography{skeleton}

\providecommand{\href}[2]{#2}\begingroup\raggedright\begin{thebibliography}{10}%
\makeatletter
\providecommand{\hrefCMSnoop }[0]{\@secondoftwo}%
\makeatother
\providecommand{\doi}{\texttt{doi:}\begingroup \urlstyle{tt}\Url}

\bibitem{TANABASHI:2018OCA}
\hrefCMSnoop {}{{Particle Data Group}, M.~Tanabashi {et~al.}, ``Review of
  particle physics'',} \textit{ Phys. Rev. D} \textbf{ 98} (2018) 030001,
\href{http://dx.doi.org/10.1103/PhysRevD.98.030001}{\doi{10.1103/PhysRevD.98.030001}}.

\bibitem{Chatrchyan:2013exa}
\hrefCMSnoop {}{{CMS} Collaboration, ``{Evidence of \cPqb jet quenching in
  \PbPb collisions at $\rootsNN =$ 2.76 \TeV}'',} \textit{ Phys. Rev. Lett}
  \textbf{ 113} (2014) 132301,
  \href{http://dx.doi.org/10.1103/PhysRevLett.113.132301}{\doi{10.1103/PhysRevLett.113.132301}},
\href{http://www.arXiv.org/abs/1312.4198}{\texttt{arXiv:1312.4198}}.

\bibitem{Baskakov:2015nxa}
A.~V. Baskakov\hrefCMSnoop {}{ {et~al.}, ``{Single top quark production in
  heavy ion collisions at energies available at the CERN Large Hadron
  Collider}'',} \textit{ Phys. Rev. C} \textbf{ 92} (2015) 044901,
  \href{http://dx.doi.org/10.1103/PhysRevC.92.044901}{\doi{10.1103/PhysRevC.92.044901}},
\href{http://www.arXiv.org/abs/1502.04875}{\texttt{arXiv:1502.04875}}.

\bibitem{Apolinario:2017sob}
\hrefCMSnoop {}{L.~Apolin\'ario, J.~G. Milhano, G.~P. Salam, and C.~A. Salgado,
  ``Probing the time structure of the quark-gluon plasma with top quarks'',}
  \textit{ Phys. Rev. Lett.} \textbf{ 120} (2018) 232301,
  \href{http://dx.doi.org/10.1103/PhysRevLett.120.232301}{\doi{10.1103/PhysRevLett.120.232301}},
\href{http://www.arXiv.org/abs/1711.03105}{\texttt{arXiv:1711.03105}}.

\bibitem{Dainese:2016gch}
A.~Dainese {et~al.}, eds., ``{Report on the physics at the HL-LHC, and
  perspectives for the HE-LHC}'', volume 7/2019 of \textit{ CERN Yellow
  Reports: Monographs}.
\newblock CERN, 2019.
\newblock
  \href{http://dx.doi.org/10.23731/CYRM-2019-007}{\doi{10.23731/CYRM-2019-007}},
ISBN~9789290835493,

\bibitem{Czakon:2011xx}
\hrefCMSnoop {}{M.~Czakon and A.~Mitov, ``{\textsc{Top++}: a program for the
  calculation of the top pair cross section at hadron colliders}'',} \textit{
  Comput. Phys. Commun.} \textbf{ 185} (2014) 2930,
  \href{http://dx.doi.org/10.1016/j.cpc.2014.06.021}{\doi{10.1016/j.cpc.2014.06.021}},
\href{http://www.arXiv.org/abs/1112.5675}{\texttt{arXiv:1112.5675}}.

\bibitem{Czakon:2013goa}
\hrefCMSnoop {}{M.~Czakon, P.~Fiedler, and A.~Mitov, ``{Total top quark pair
  production cross section at hadron colliders through O($\alpha_S^4$)}'',}
  \textit{ Phys. Rev. Lett.} \textbf{ 110} (2013) 252004,
  \href{http://dx.doi.org/10.1103/PhysRevLett.110.252004}{\doi{10.1103/PhysRevLett.110.252004}},
\href{http://www.arXiv.org/abs/1303.6254}{\texttt{arXiv:1303.6254}}.

\bibitem{Catani:2019hip}
S.~Catani\hrefCMSnoop {}{ {et~al.}, ``{Top quark pair production at the LHC:
  Fully differential QCD predictions at NNLO}'',} \textit{ JHEP} \textbf{ 07}
  (2019) 100,
  \href{http://dx.doi.org/10.1007/JHEP07(2019)100}{\doi{10.1007/JHEP07(2019)100}},
\href{http://www.arXiv.org/abs/1906.06535}{\texttt{arXiv:1906.06535}}.

\bibitem{Sirunyan:2017ule}
\hrefCMSnoop {}{{CMS} Collaboration, ``Measurement of the inclusive $\ttbar$
  cross section in pp collisions at {$\roots =$ 5.02 \TeV} using final states
  with at least one charged lepton'',} \textit{ JHEP} \textbf{ 03} (2018) 115,
  \href{http://dx.doi.org/10.1007/JHEP03(2018)115}{\doi{10.1007/JHEP03(2018)115}},
\href{http://www.arXiv.org/abs/1711.03143}{\texttt{arXiv:1711.03143}}.

\bibitem{Sirunyan:2017xku}
\hrefCMSnoop {}{{CMS} Collaboration, ``{Observation of top quark production in
  proton-nucleus collisions}'',} \textit{ Phys. Rev. Lett.} \textbf{ 119}
  (2017) 242001,
  \href{http://dx.doi.org/10.1103/PhysRevLett.119.242001}{\doi{10.1103/PhysRevLett.119.242001}},
\href{http://www.arXiv.org/abs/1709.07411}{\texttt{arXiv:1709.07411}}.

\bibitem{Chatrchyan:2008zzk}
\hrefCMSnoop {}{{CMS} Collaboration, ``The {CMS} experiment at the {CERN}
  {LHC}'',} \textit{ JINST} \textbf{ 3} (2008) S08004,
\href{http://dx.doi.org/10.1088/1748-0221/3/08/S08004}{\doi{10.1088/1748-0221/3/08/S08004}}.

\bibitem{CMS-PAS-LUM-17-002}
\href {http://cds.cern.ch/record/2628652}{{CMS Collaboration}, ``{CMS}
  luminosity measurement using 2016 proton-nucleus collisions at
  nucleon-nucleon center-of-mass energy of 8.16 {TeV}'',} CMS Physics Analysis
  Summary CMS-PAS-LUM-17-002, 2018.

\bibitem{Sirunyan:2020kov}
\hrefCMSnoop {}{{CMS} Collaboration, ``{Evidence for top quark production in
  nucleus-nucleus collisions}'',} (2020).
  \href{http://www.arXiv.org/abs/2006.11110}{\texttt{arXiv:2006.11110}}.
Submitted to \textit{Phys. Rev. Lett.}

\bibitem{Alwall:2014hca}
J.~Alwall\hrefCMSnoop {}{ {et~al.}, ``The automated computation of tree-level
  and next-to-leading order differential cross sections, and their matching to
  parton shower simulations'',} \textit{ JHEP} \textbf{ 07} (2014) 079,
  \href{http://dx.doi.org/10.1007/JHEP07(2014)079}{\doi{10.1007/JHEP07(2014)079}},
\href{http://www.arXiv.org/abs/1405.0301}{\texttt{arXiv:1405.0301}}.

\bibitem{ESKOLA:2016OHT}
\hrefCMSnoop {}{K.~J. Eskola, P.~Paakkinen, H.~Paukkunen, and C.~A. Salgado,
  ``{EPPS16: Nuclear parton distributions with LHC data}'',} \textit{ Eur.
  Phys. J. C} \textbf{ 77} (2017) 163,
  \href{http://dx.doi.org/10.1140/epjc/s10052-017-4725-9}{\doi{10.1140/epjc/s10052-017-4725-9}},
\href{http://www.arXiv.org/abs/1612.05741}{\texttt{arXiv:1612.05741}}.

\bibitem{Alioli:2010xd}
\hrefCMSnoop {}{S.~Alioli, P.~Nason, C.~Oleari, and E.~Re, ``A general
  framework for implementing {NLO} calculations in shower {Monte Carlo}
  programs: the {\POWHEG} {BOX}'',} \textit{ JHEP} \textbf{ 06} (2010) 043,
  \href{http://dx.doi.org/10.1007/JHEP06(2010)043}{\doi{10.1007/JHEP06(2010)043}},
\href{http://www.arXiv.org/abs/1002.2581}{\texttt{arXiv:1002.2581}}.

\end{thebibliography}\endgroup

\end{document}